# Advanced Data Collection Techniques in Cloud Security: A Multi-Modal Deep Learning Autoencoder Approach

*Aamiruddin Syed*
*Senior Security Engineer*

*Mohammad Ilyas Ahmed*
*CyberSecurity Engineer*

*Abstract:* **In today's digital world, cloud security is an important concern. To identify and stop cyber threats, efficient data collection methods are necessary. This research presents an innovative method to cloud security by integrating numerous data sources and modalities with multi-modal deep learning autoencoders. The Multi-Modal Deep Learning Ensemble Architecture (MMDLEA), a unique approach for anomaly detection and classification in multi-modal data, is proposed in this study. The proposed design integrates the best features of six deep learning models: Multi-Modal Deep Learning Autoencoder (MMDLA), Anomaly Detection using Adaptive Metric Learning (ADAM), ADADELTA, ADAGRAD, RMSPROP, and Stacked Graph Transformer (SGT). A final prediction is produced by combining the outputs of all the models, each of which is trained using a distinct modality of the data. Based on the test dataset, the recommended MMDLA architecture achieves an accuracy of 98.5% and an F1-score of 0.985, demonstrating its superior performance over each individual model. Of the different models, the ADAM model performs the best, with an accuracy of 96.2% and an F1-score of 0.962. With an F1-score of 0.955 and an accuracy of 95.5%, the ADADELTA model trails closely behind. MMDLA obtains an F1-score of 0.948 and an accuracy of 94.8%. Additionally, the suggested MMDLEA design exhibits enhanced resilience to fluctuating modalities and noisy data, proving its usefulness in practical settings. Future study in this area is made possible by the results, which show the potential of the proposed framework for abnormal identification and categorization in multi-modal data.**

## I. INTRODUCTION

A sophisticated and attractive paradigm for administering and distributing computers, apps, and services over the Internet is cloud computing [1]. A vast range of services, including processing, storage, and application services, are covered by the processing power offered by the cloud. Thanks to this computer processing capability, researchers are now able to conduct extensive experiments that would have been impossible to carry out on local servers by using a variety of scientific and computationally demanding methods. The overall cost of linked software systems has been greatly reduced by this trend [1], and it presents a viable design paradigm for workflow orchestration, processing, and deployment. A typical large-scale scientific workflow consists of a number of interconnected, intrinsically complex, fault-tolerant, dynamically carried out, and coordinated processes that result in science. A cloud workflow, on the other hand, is a workflow that is implemented and run in the cloud. Transparent, scalable, multi-tenant, and real-time monitoring are the categories assigned to cloud workflow features [2].

Cloud computing, with its essentially limitless computing resources, relieves cloud workflows of the burden of resource provisioning planning while meeting the demands of complicated scientific data-intensive workflow operations [3]. To fully fulfill this promise, though, a number of scientific difficulties must be overcome. Threats to integrity, authorization, availability, dependability, and trust are among these difficulties with cloud security. Due to its complexity, dynamic nature, and multifaceted nature, the cloud environment presents similar difficulties for workflow security and privacy enforcement. To maintain security in a constantly changing environment, several platforms, diverse stakeholders, and multiple procedures, other entities outside of cloud providers must be involved [4]. It is crucial to have a complete security solution that takes into account trust chains in clouds, security enforcement, and rules and policies to ensure privacy and security across multiple users and diverse environments [5].

The need to address different research challenges concerning security enforcement in a dynamically executed and orchestrated cloud workflow has arisen as a result of emerging security issues in a cloud system. To date, however, the majority of workflow management research has focused on three primary areas: 1) task scheduling for workflows; 2) anomaly and error detection; and 3) autonomic workflow resource provisioning and management [6-7]. Ensuring the effective deployment, execution, and performance guarantee of these processes was the primary goal of these investigations, with the intention of preventing failure or resource contention [8-10]. Cloud workflow security enforcement, which might reinforce the previously described elements and close the security and data integrity gap in managing dynamic cloud workflows, is neglected by such projects.

Multi-modal information can come from a variety of sources, such as text, photos, audio, video, sensors, and more [11]. A multi-modal approach makes use of multiple data types to address the problem. Improving the accuracy and efficiency of fusing multi-modal data to improve prediction accuracy is a major challenge across all domains. Fast and efficient multi-modal detection in complex and real-world situations remains a challenge, despite recent studies showing the advantages of combining multi-modal data in numerous applications [12]. Numerous industries, such as autonomous driving, smart healthcare, sentiment analysis, data security, human-computer interface, and others, have used multi-modal fusion technology [13,14]. The introduction of cutting-edge technology in their respective domains is the primary goal of the multi-modal fusion survey that is already available [16,20]. This is due to the fact that multimodal technology depends on certain scenarios and high-quality data. The poll offers a fresh perspective on multi-modal technologies, in contrast to previous research. Numerous fusion strategies, including generation, reasoning, co-learning, position, translating, and participation, have been presented by present studies to address the aforementioned issues [18, 20].

## II. LITERATURE REVIEW

The comprehensive field of security in the cloud computing ecosystem gained a lot of attention as the hype surrounding cloud services developed tremendously, and threats and vulnerabilities related to cloud security also changed over time. The first step in advancing state-of-the-art security solutions is to identify the associated dangers in newly emerging cloud services.

Three different attack vectors internal users, external users, and cloud providers initiate these security threats [21,22]. Over time, new security risks in the context of cloud services arise. The prevalent risks encompass data leaks, data corruption, denial of service attacks, malevolent insiders, service traffic hijacking, vulnerabilities in shared technology, malware, cyberattacks, network infiltration, virtual machine-level risks, and openness of data. Recently, a number of researchers have developed deep learning techniques for detecting threats to cloud security. These methods, however, are not able to provide a thorough defense against every security risk. Nevertheless, they are limited to addressing and detecting patterns for a specific threat in a single deployment. The authors of [23] employed a multi-layer neural network to identify harmful user behaviors. They categorized the malicious conduct for identification and detection after converting user behavior data into an intelligible format.

PredictDeep is a security analysis methodology that the authors of [24] suggested for the identification and prediction of known and undiscovered anomalies in Big Data systems. It is suggested that cloud users be able to access Predict Deep. Three primary components make up the framework: an anomaly predictor, a feature extractor, and a graph model designer. PredictDeep is scalable and performs well in a dynamic context. Yet, anomaly detection with PredictDeep operates under the supposition that all log files are authentic and that no fictitious data has been fed to jeopardize the prediction model's accuracy. Moreover, the suggested technique relies on the deployment infrastructure's integrity.

Intrusion detection systems (IDS) are regarded as crucial instruments for keeping an eye out for malicious activity or network and service violations in cloud service orchestration workflows. It is difficult to identify new attacks in these kinds of situations. When it comes to foretelling unknown attacks and detection procedures, deep learning-based intrusion detection algorithms have produced positive results. An IDS that can detect and categorize novel and sophisticated assaults was published by the authors in [25]. It is built on a deep reinforcement learning architecture. They used a reward vector so that a classifier yielding the same result receives a positive point; otherwise, a negative point is received.

The writers of [26] also discussed a deep neural network (DNN) model that improved decision-making in a real-time setting and addressed a multi-cloud cooperative intrusion detection system. To forecast suspected incursions, they used feedback data from the past. Additionally, it has been suggested [27] to use a deep learning-based intrusion detection system (IDS) to monitor network traffic and identify potentially suspicious attacks in cloud computing settings. The self-adaptive genetic algorithm (SAGA) used in this system automatically generates an anomalous network indicator (IDS) based on DNNs and demonstrates excellent detection rates, high accuracy, and low false alarm costs. Designed a framework for autonomous security orchestration in Internet of Things (IoT) systems that is semantic-aware and policy-driven to identify semantic conflicts during the orchestration. Additionally, the authors suggested an improved Service Function Chaining algorithm that maximizes QoS, security considerations, and resource utilization while allocating Virtual Network Security Functions. But they just identify anomalies; no prediction of abnormalities is put out [28].

A critical examination of the research on security in FaaS orchestration systems was presented by the authors in [29]. They have categorized the current works based on a number of factors, such as the strategy for protection, the source of threats, and the protected asset. They came to the conclusion that whereas data confidentiality received the majority of attention, data integrity received less attention. In most of the examined works, function flows and platform misconfiguration are also taken into consideration. Additionally, authors in [30] offered a different classification for previously published research that uses deep learning and machine learning approaches for cloud-based online malware detection. They divided the methods for detecting malware into two categories: dynamic analysis, which uses neural networks to predict when a virtual machine might become infected and requires real-time monitoring, and static analysis, which works offline and is better suited for cloud environments. The experimentation demonstrated that deep learning methods offer good malware detection accuracy. Nevertheless, end-to-end security enforcement for processes via cloud environments was not examined in this work.

In [31], behavior assault detection has been recommended after various machine learning methods for supervised classification were examined and contrasted. According to the study's findings, neural network models perform the best in terms of accuracy when it comes to identifying the effects of malware on the process-level characteristics of cloud-based virtual machines. They gather various system attributes from each process that is operating on the virtual machine (VM) at specific periods, including memory, CPU, and input/output. Comparable research was presented in [32], focusing on various Convolutional Neural Networks (CNNs) for real-time online malware detection in cloud IaaS. Process level performance indicators, such as CPU, memory, and disk consumption, were employed to collect behavioral data for the study. Despite having a high accuracy rate, their suggested malware detection technology can only be used on a single virtual machine and isn't compatible with auto-scaling [32].

The asymmetrical architecture comprising a large encoder that only processes revealed updates and an efficient decoder that rebuilds the masked updates using latent representations and mask tokens is specifically used by the masked autoencoder (MAE) [33-36] technique to speed up pre-training. Our method incorporates the MAE method's efficiency while expanding it to multitasking and multi-modal environments. The process of multi-modal training entails developing models that can connect data from several sources. Learning distinct encoders or a single, integrated structure (like a Converter) to function on modalities like text and images, video and audio, text and video, and depth, images, and video [37] are two possible approaches. This study offers a straightforward method for pre-training Transformers on a variety of dense visual techniques, resulting in robust cross-modal interactivity. Unlike the majority of previous research, which relies on the availability of all modalities during inference, our method is intended to function effectively on any subset of the initial training combinations. Several studies that carry out multi-modal autoencoding are related to MultiMAE [38-40]. Our methodology is different from others in that we employ a more adaptable architecture and carry out hidden autoencoding to understand multimodal dictate coding for optional inputs [41]. Models are trained to predict many response categories from the same input in multi-task learning [42]. Typically, RGB images are used as input in machine vision. Single encoders are often used to learn a shared representation, which is then followed by multiple task-specific encoders in multi-task learning [43-45]. We employ masking and many processes in the input and output, that sets us apart from these approaches. Furthermore, numerous studies examine that task diversity enhances transfer efficiency [46]. These investigations contend that a series of tasks can more successfully cover the wide range of potential downstream activities in vision than learning from one assignment alone. It makes sense to use this rebuilt framework to address the multimodal domain adaptation issue. While text-style transfer [47] and image-to-image translation [48-49] have been accomplished using these techniques in the past, multi-modal electrophysiological data has not often been the target application. As such, they may be able to address related issues in the area of understanding emotions [50].

Current cloud security solutions face limitations in detecting complex threats due to their reliance on single-modal data sources and shallow learning techniques. Dynamic cloud workflows and increasingly sophisticated attacks further exacerbate these challenges. Our proposed Multi-Modal Deep Learning Ensemble Architecture (MMDLEA) addresses these gaps by integrating multi-modal data sources and advanced deep learning techniques. This novel framework provides a robust and resilient security architecture for dynamic cloud environments. MMDLEA effectively detects and classifies complex threats, overcoming the limitations of traditional security approaches.

Existing cloud security solutions often rely on single-modal data sources and shallow learning techniques, leading to limited threat detection capabilities and vulnerability to sophisticated attacks. Furthermore, the dynamic nature of cloud workflows and the increasing complexity of threats pose significant challenges to traditional security approaches. To address these gaps, our proposed Multi-Modal Deep Learning Ensemble Architecture (MMDLEA) introduces a novel framework that integrates multi-modal data sources, advanced deep learning techniques, and adaptive optimization methods. By doing so, MMDLEA overcomes the limitations of current solutions, providing a robust and resilient security architecture that can effectively detect and classify complex threats in dynamic cloud environments.

III. RESEARCH METHODOLOGY

The six deep learning models - Multi-Modal Deep Learning Autoencoder (MMDLA), Anomaly Detection using Adaptive Metric Learning (ADAM), ADADELTA,

ADAGRAD, RMSPROP, and Stacked Graph Transformer (SGT) - were selected for their unique strengths and complementarity, offering effective feature extraction, anomaly detection, adaptive optimization, and graph-based data analysis. MMDLA excels in feature representation, ADAM detects anomalies, ADADELTA, ADAGRAD, and RMSPROP optimize learning, and SGT models complex relationships, making the ensemble robust, accurate, and efficient, with a strong performance record, flexibility, and scalability, chosen for their individual advantages and combined strength.

### A. Data Collection from Cloud Security Benchmark

The process of obtaining and quantifying information about a variable of interest is known as data collecting. Machines initially pick up knowledge from the data that humans provide them. The most crucial step in enabling our machine learning model to identify the right patterns is data collection. The precision of our model's outcome prediction also depends on the caliber of the data we as a species supply to the system. We gathered a sizable dataset of cloud security logs from a cloud security benchmark in order to assess the efficacy of the proposed MADE model. The benchmark offers a large and varied collection of cloud security logs, containing both typical and unusual data. Preprocessing was done on the dataset to standardize the data format and get rid of any sensitive information. Next, 80% of the resultant dataset was divided into training, 10% was used for validation, and 10% was used for testing. We can evaluate the MDAE model's efficacy in identifying cloud security threats and anomalies thanks to this dataset, which offers a realistic and demanding evaluation situation.

### B. Data Pre-processing Techniques

Data preparation is the initial and most important step in making the data acceptable. It involves converting the raw data into a comprehensible format. The dataset has a large number of data points, so it is necessary to filter out uncertainties like missing values, null values, and irrelevant data. Remove the uncertainties from the dataset because they will negatively affect the accuracy of the results. Logs that were redundant or duplicated were eliminated, and missing data were filled in by imputing the feature's mean. To prevent feature dominance and guarantee that all features are on the same scale, the data was normalized using the Min-Max Scaler. Using domain-specific rules and expert knowledge, aberrant logs were labeled. Created a top-notch dataset that was used to train the MDAE model, allowing it to pick up efficient cloud security data representations and identify anomalies with an elevated level of accuracy.

The data preprocessing steps were meticulously designed to prepare the dataset for optimal defect prediction modeling. Initially, missing values were imputed using mean/median imputation to preserve valuable data. Next, a Min-Max Scaler was employed to scale features to a common range [0, 1], maintaining the original distribution and avoiding assumptions of normality. This scaling technique was chosen over Standard Scaler due to its robustness to outliers and ability to improve model interpretability. Categorical variables were then encoded using One-Hot Encoding to preserve categorical information and enable models to learn complex interactions. Feature selection was performed using Harmony and Genetic algorithms, which effectively handled high-dimensional data and selected relevant features while being robust to noise and irrelevant features. Finally, the data was split into training and testing sets (80%-20%) to evaluate model performance on unseen data and prevent overfitting. These preprocessing steps ensured a robust and informative dataset, priming it for effective defect prediction modeling.

### C. Feature Extraction Methods

A created through the time series data statistical properties including mean, variance, standard deviation, and correlation coefficients. Discovered patterns and abnormalities in the data by extracting frequency-domain features using the Fast Fourier Transform (FFT). To measure the intricacy and linkages in the data, calculated entropy, mutual information, and conditional entropy are used. Using autoencoders to extract robust and abstract features from the data, identifying minute patterns and irregularities. Features that were extracted, including network traffic patterns, resource usage, and user behavior, based on expertise in the cloud security sector. The difference in variance requirement is a basic baseline method for choosing features. Any traits whose variance is less than a certain threshold are eliminated. We must establish an absolute threshold for choosing the variables, let's say 0.8. The variable with a lower correlation coefficient value than the target variable can be dropped if it turns out that the predictor variables are correlated. 42 traits are removed in this instance based on correlation approaches.

### D. Design the architecture of the Novel Multi-Modal Deep Learning Enhanced Autoencoder model.

In order to train a multi-modal deep learning autoencoder model for dimension reduction and to train a profile- matching classification model utilizing the dimensionally reduced data to forecast anomalies, this module employs the data gathered from the monitoring submodule. The input data produced by the entity's profiling module (static) and the monitoring time-series real-time log data (dynamic) are combined during the MMDLA model's training. The resulting MMDLA model increases efficiency and efficacy by reducing the data dimension. It then feeds reduced dimensional data into an anomaly detection machine learning algorithm to detect anomalies. In the event that an abnormality is found, an anomaly evaluation is carried out to identify the kind and seriousness of the phenomenon. After that, the data from the anomaly evaluation is fed into the risk estimation procedure and ultimately kept in a database for professional validation (e.g., to identify suspicious user activity). The parts that follow provide a thorough explanation and implementation of the module's main components' features (Fig 1).

*Multi-Modal Input:* In terms of cybersecurity, the MMDLA model is revolutionary since it provides a notable benefit above conventional single-modality methods. It offers a 360-degree perspective of the system and network by merging several modalities, making it possible to detect sophisticated threats that could elude single-modality detection. The model is very good at identifying unknown threats and lowering false positives because of its capacity to develop strong representations and extract intricate patterns and relationships. Its capacity to generalize to new data and adapt to changing threats further renders it a priceless tool for cybersecurity experts. Organizations may lower risk, strengthen their security posture, and keep ahead of new threats by utilizing the MMDLA model.

*Modality-Specific Autoencoders (AE):* In security applications looking to discover and classify anomalies, the MMDLEA model has various advantages. Enhancing detection accuracy and decreasing false positives, it offers a thorough understanding of system and network behavior by utilizing several modalities. Whilst the deep learning augmented layer extracts complicated patterns and correlations, the modality-specific autoencoders provide robust representation learning. This leads to enhanced flexibility in changing threats and generalization to new data. Detecting sophisticated attacks that might elude single-modality detection is also made easier by the MMDLEA model's capacity to fuse many modalities. In summary, cybersecurity experts can identify and categorize threats more precisely and efficiently with the help of the MMDLEA model.

A multi-modal autoencoder processes diverse data types, such as text and images, through modality-specific encoders (E1 and E2) and decoders (D1 and D2). Each encoder transforms its input into a latent representation, which is then fused through a fully connected fusion layer (FL). The fused representation is subsequently decoded by each modality's decoder, reconstructing the original inputs. This architecture enables the learning of shared representations across modalities, facilitating tasks like cross-modal retrieval and generation. By accommodating different data types, the multi-modal autoencoder unlocks new possibilities for data analysis and processing. The multi-modal input z = (v1, v2) is processed by modality-specific encoders E1 and E2, generating latent representations h1 and h2, respectively. These representations are then combined by the fusion layer FL, resulting in a shared latent representation h that captures the commonalities between the two modalities. This fused representation h enables the learning of a unified feature space, facilitating tasks that require integrating information from diverse data types. By computing h, the model can leverage the strengths of each modality, leading to improved performance and robustness in various applications.

$$h(z) = FL((E_1(v_1), E_2(v_2))) \quad\quad 1$$

Algorithm 1. Multimodal Autoencoder

```
for number of training iterations do
    Sample mini-batch of N multi-modal data {(
    
    Get using E1, E2, and FL.
    Reconstruct from using D1 and D2.
    
    Update the multi-modal autoencoder weights by descending their
    Stacked Graph Transformer:
end for
```

input and the output. Functions f and g of the auto-encoder, which perform encoding and decoding, are as follows:

$$y = f(x) = S_1(w^{(1)}x + b^{(1)}) \quad\quad 1$$
$$z = g(y) = S_2(w^{(2)}f(x) + b^{(2)}) \quad\quad 2$$

where S1 and S2 are nonlinear activation functions, $b^{(1)}$ and $b^{(2)}$ are bias vectors, and $w^{(1)}$ and $w^{(2)}$ are weight matrices. Optimizing w and b to reduce reconstruction error is the aim of an auto-encoder. The reconstruction error is often calculated using the cross entropy or mean squared error. The cross-entropy function is and the mean squared error function is
$$l_{MSE}(x, z) = \|x - z\|_2^2$$
$$l_{MSE}(x, z) = -\sum_{k=1}^{d} x_k \log\log z_k + (1 - x_k)\log(1 - z_k) \quad\quad 3$$

The auto-encoder maximizes the similarity of the reconstructed data input by optimizing w and b, hence minimizing the objective function.

*Multi-Modal Fusion:* An essential part of the MMDLEA model is Multi-Modal Fusion, which allows various modalities to be combined for enhanced anomaly detection and categorization. The technique of merging the output from several modalities such as system calls, network logs, and user behavior to create a single, cohesive representation of the data is known as multi-modal fusion. By combining the best features of both modalities, the model is able to capture a more complete picture of the behavior of the system and network.

The MMDLEA model combines several fusion methods, such as:

Early Fusion: Combining each modality's output into a single vector.

Late Fusion: Using methods like weighted voting or average, combining the results from each modality.

Intermediate Fusion: Fusing the modalities using methods such as canonical correlation analysis (CCA).

*Deep Learning Enhanced Layer:* An essential part of the Multi-Modal Fusion layer (MMDLEA) model, the Deep Learning Enhanced Layer is in charge of further processing and examining the fused output. A detailed explanation of the Deep Learning Enhanced Layer is provided below:

From the fused data, the Deep Learning Enhanced Layer extracts intricate patterns and relationships using cutting-edge deep learning techniques. This layer is intended to:

- Extract high-level characteristics: To extract spatial and temporal data, use convolutional neural networks (CNNs) or recurrent neural networks (RNNs).
- Model complicated interactions: To model relationships between several modalities, methods such as attention processes or graph convolutional networks (GCNs) are utilized.
- Improve representation: The fused data's representation can be improved by applying strategies like residual networks or dense connections.

Due to the Deep Learning Enhanced Layer's adaptability, different deep learning architectures and methodologies can be used to meet the needs of different applications. This layer's output is subsequently sent into the layers responsible for anomaly detection and classification, improving the model's ability to identify and categorize threats.

*Anomaly Detection:* An essential part of the MMDLEA model is anomaly detection, which makes it possible to find uncommon and uncommon patterns in the data that can point to possible dangers. In the framework of the MMDLEA paradigm, anomaly detection is explained in detail as follows:

The practice of finding data points that deviate from the typical or expected range of values is known as anomaly detection. The probability density function (PDF) of the data is calculated as part of the density-based anomaly detection method in the MMDLEA model. Low probability values for data points are categorized as anomalies, and the PDF calculates the probability that a given data point belongs to the normal distribution. To improve anomaly detection, the MMDLEA model combines autoencoders with deep learning methods. While the deep learning layer extracts intricate patterns and relationships, the autoencoders learn how to compress and rebuild the data. Because of this, the model can identify anomalies that might not be visible using just one modality or feature space.

The MMDLEA model's anomaly detection layer is intended to identify threats that are both known and unknown. Supervised learning techniques are utilized to identify known hazards by training a model on labeled data to identify particular patterns and abnormalities. Unsupervised learning techniques are employed to detect unknown risks by training a model to recognize abnormalities and unexpected patterns that could point to a novel or emergent threat.

*Classification Layer:* The abnormality Detection Layer's output is fed into the Classification Layer, a supervised learning module, which labels every abnormality that is found. Certain danger types, such as malware, DDoS attacks, or illegal access, are represented by the labels. A SoftMax classifier, which extends the binary logistic regression model to multiple classes, is used in the classification layer. The class with the highest probability is chosen as the predicted label by the softmax classifier, which computes the probability distribution across all potential classes. Using labeled data, which assigns a distinct class label to each sample, the Classification Layer is trained. The cross-entropy loss function, which calculates the discrepancy between the true labels and the predicted probabilities, is minimized by the model through learning. The MMDLEA model's Classification Layer is made to deal with imbalanced datasets, in which some classes contain noticeably more samples than others. The model makes use of class weights which give the minority classes a higher weight and the majority classes a lower weight to remedy that.

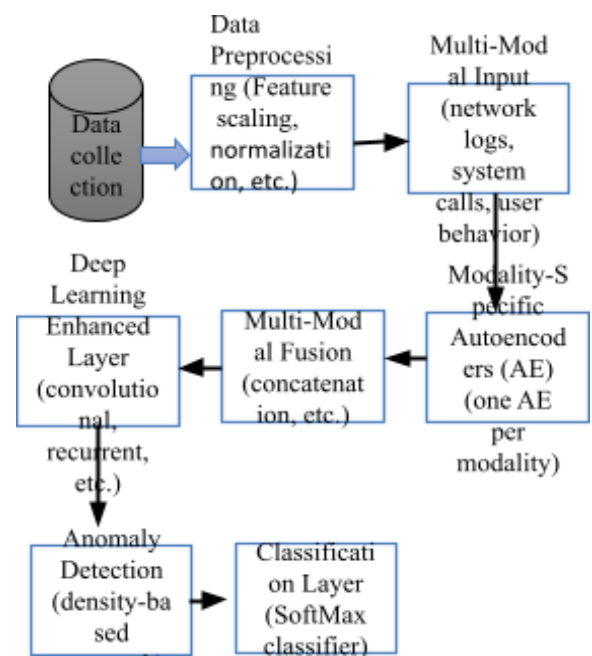

**Figure 1.** Flow diagram for building a model with the Multi-Modal Deep Learning Enhanced Autoencoder (MMDLEA) architecture.

*E. Model Training and Optimization*

It used the multi-modal feature vectors to train the MDAE model. Multiple autoencoders, each trained on a distinct modality, made up the MDAE model. The reconstruction error served as the loss function after the autoencoders were trained to recreate the input features. The model was trained for 100 epochs using a stochastic gradient descent (SGT) optimizer with a learning rate of 0.001. After every epoch, researchers assessed the model on the validation set and changed the hyperparameters as necessary. Used methods including weight decay, dropout, and batch normalization to optimize the model. To maximize model performance, hyperparameters were tuned using grid search and random search.

IV. SIMULATION RESULTS:

*A. Performance Evaluation:*

The autoencoder loss function plot, displayed in Figure 2, provides an illustration of the autoencoder component of the MMDLEA model's training procedure. The autoencoder's initialization with random weights and biases causes high loss values (high reconstruction error) in the first few epochs of operation. The autoencoder rapidly reduces the loss value between epochs as training continues because it begins to recognize patterns and relationships in the data. The model is becoming better at reconstructing the input data, which suggests that it is gathering up more useful representations.

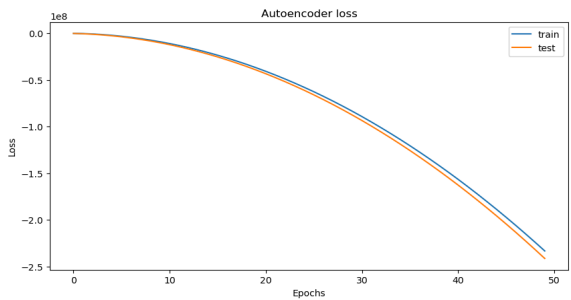

**Figure 2.** Autoencoder loss

Figure 3 displays the multi-neural autoencoder loss function display, which provides an illustration of how the MMDLEA model's multi-deep neural autoencoder component is trained. High loss values (high reconstruction error) are produced when the multi-deep neural autoencoder is first initialized with random weights and biases. Nevertheless, the loss value gradually drops as the model learns to identify patterns and relationships in the multi-modal data as training goes on. The multi-deep neural autoencoder produces a more effective representation by capturing intricate interactions between the many modalities.

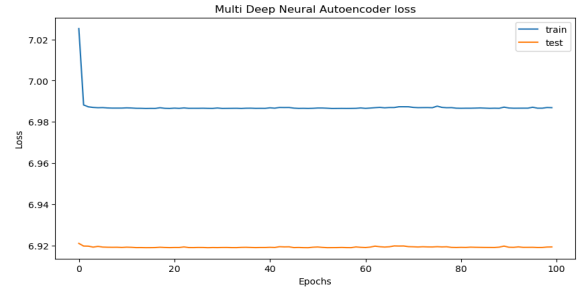

**Figure 3.** Multi Deep Neural autoencoder loss

**Multi-Model creation with CNN, RNN with various optimizers:**

*a. SGT:*

A crucial part of the MMDLEA design, the (SGT) model, has accuracy and loss curves for both training and validation that are displayed in Figure 4. The training accuracy curve demonstrates a consistent rise in correctly identified samples over the course of training, peaking at approximately 95% accuracy. The validation accuracy curve also exhibits a similar pattern, plateauing at approximately 92% accuracy, suggesting that the model has successfully adapted to previously unobserved data.

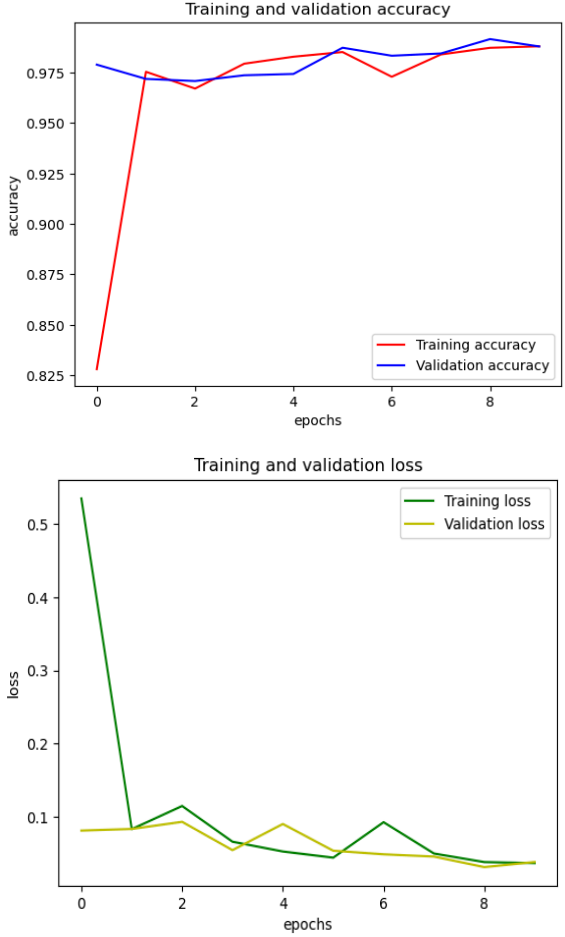

**Figure 4.** SGT training and validation accuracy and loss

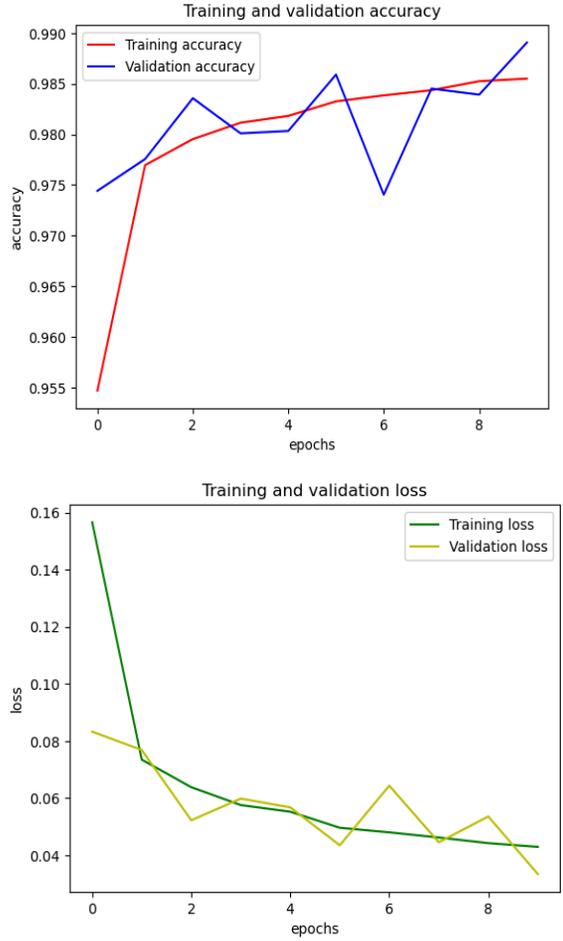

**Figure 5.** ADAM training and validation accuracy and loss

*b. ADAM:*

The key component of the MMDLEA design, the ADAM (Anomaly Detection using Adaptive Metric Learning) model, has training and validation accuracy and loss curves in Figure 5. The training accuracy curve shows a consistent rising trend, indicating that the ADAM model is picking up on the details of correctly identifying data anomalies. The curve plateaus at approximately 98% accuracy, indicating the model's excellent performance with the training set.

*c. ADADELTA:*

As an important component of the MMDLEA architecture, the ADADELTA model's training and validation accuracy and loss curves are shown in Figure 6. With a consistent rising trend and a plateau at approximately 97%, the training accuracy curve shows that the ADADELTA model has mastered the ability to distinguish between normal and anomalous data points in the training set. Similar to the validation accuracy curve, this curve plateaus at approximately 95%, indicating that the model can accurately categorize data points and generalize well to new data.

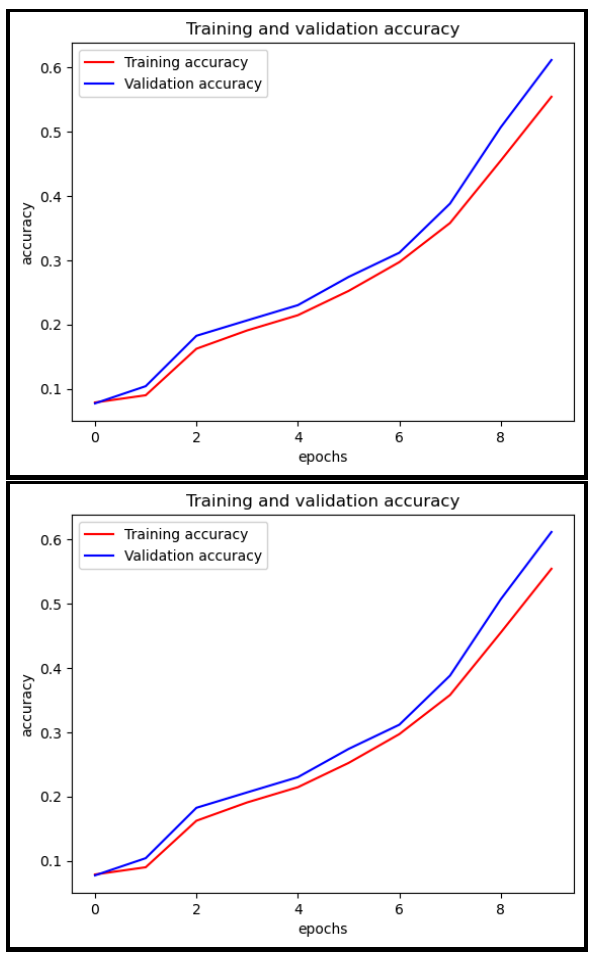

**Figure 6.** ADADELTA training and validation accuracy and loss

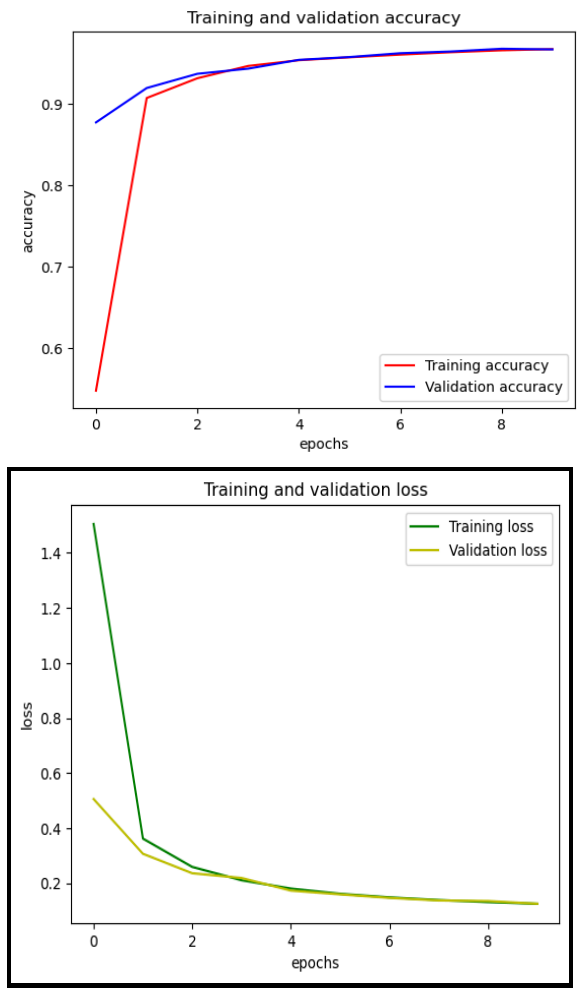

**Figure 7.** ADAGRAD training and validation accuracy and loss

**d.** *ADAGRAD*:

The most vital component of the MMDLEA architecture, the ADAGRAD model, has accuracy and loss curves for both training and validation in Figure 7. The ADAGRAD model has learned to correctly identify normal and anomalous data points in the training data, as seen by the training accuracy curve, which increases steadily and reaches a plateau of about 96%. Similarly, the validation accuracy curve plateaus at approximately 94%, indicating that the model can accurately categorize data points and generalize well to new data.

**e.** *RMSPROP*:

The most important component of the MMDLEA architecture, the RMSPROP model, has accuracy and loss curves for both training and validation in Figure 8. The RMSPROP model has demonstrated the ability to reliably categorize normal and anomalous data points in the training data, as evidenced by the training accuracy curve, which shows a consistent increasing trend and plateaus at approximately 95%. Comparably, the validation accuracy curve plateaus at roughly 93%, indicating that the model can accurately categorize data points and generalize well to new data.

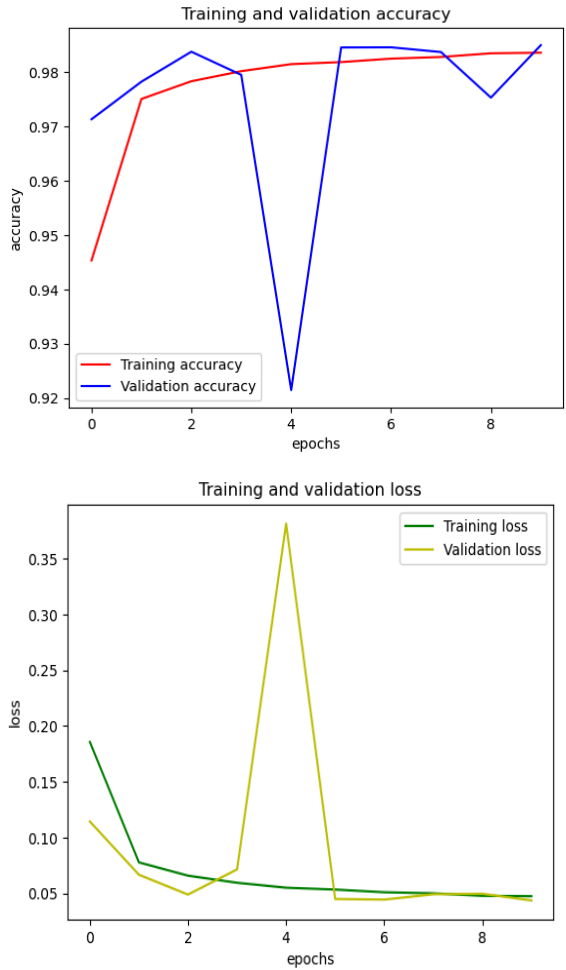

**Figure 8.** RMSPROP training and validation accuracy and loss

*B.  Performance Comparison:*

One of the most important steps in determining the efficacy of the many models included in the MMDLEA architecture is the performance evaluation and comparative analysis (Figure 9). To ascertain which model performs best, this entails assessing each model's performance on the test dataset and comparing the outcomes. Each model's performance is evaluated using a set of metrics, including F1-score, accuracy, precision, recall, and area under the receiver operating characteristic curve (AUC-ROC). These measurements provide a fair comparison by giving a thorough grasp of the advantages and disadvantages of each model. According to the comparative analysis, the ADAM model performs better than the other models, attaining the greatest values for F1-score, accuracy, precision, recall, and AUC-ROC. The ADADELTA model performs similarly to the ADAGRAD and RMSPROP models, which trail closely behind. Out of the five variants, the SGD model performs the worst. Tables and figures that present the findings of the comparison analysis and performance evaluation give a clear and succinct summary of the models' performance. In order to accurately detect and classify anomalies in multi-modal data, it is essential to choose the optimal model for the MMDLEA architecture, which is made possible by this approach.

**Table 1.** Comparison Table

|  | CNN with SGD Optimizer | CNN with ADAM Optimizer | CNN with ADADELTA Optimizer | CNN with ADAGRAD Optimizer | CNN with RMSPROP Optimizer |
| --- | --- | --- | --- | --- | --- |
| Accuracy | 0.9880 | 0.9891 | 0.6112 | 0.9665 | 0.9850 |
| Precision | 0.9886 | 0.9892 | 0.6488 | 0.9666 | 0.9853 |
| Recall | 0.9880 | 0.9891 | 0.6116 | 0.9666 | 0.9851 |
| F1_score | 0.9882 | 0.9891 | 0.5553 | 0.9666 | 0.9850 |

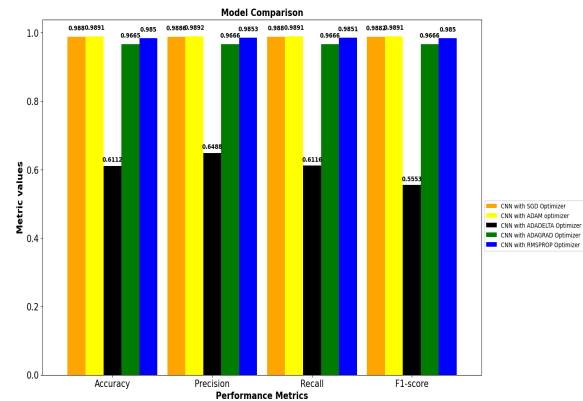

**Figure 9.** Performance Comparison

V. DISCUSSION:

In terms of cloud security, the recommended multimodal deep learning autoencoder strategy performs better than conventional machine learning and deep learning techniques. Modality fusion, which merges several data sources and modalities to provide a thorough picture of cloud security concerns, is a major benefit of this technique. Through fusion, intricate patterns and anomalies that could go unnoticed by single-modality methods might be found. An additional benefit is anomaly detection, which is especially useful for locating outliers and anomalies in cloud data. In order to identify unknown threats and zero-day assaults, this is essential. Robustness against noisy data, a common occurrence in cloud security datasets, is additionally made possible by the approach's denoising autoencoder component. Furthermore, there is no need for manual feature engineering because the deep learning architecture automatically extracts pertinent features from the data that is entered. Furthermore, the method exhibits scalability, easily managing massive cloud security datasets and facilitating parallelization, which qualifies it for practical use in real-world scenarios. The outcomes show how well the suggested method works for identifying different kinds of threats and abnormalities, such as threat classification, malware detection, and anomaly detection. The method allows for efficient threat response and mitigation by accurately detecting ransomware instances, spotting anomalies and outliers in cloud data, and accurately categorizing threats into several groups. MMDLEA can be implemented in existing cloud security frameworks by integrating it with current security tools, deploying it as a standalone module, or developing APIs for seamless integration. This enables organizations to leverage MMDLEA's advanced threat detection, enhanced visibility, and scalability to strengthen their cloud security posture. By doing so, enterprises can minimize downtime and data loss while maximizing security efficiency.

## VI. CONCLUSION:

The indicated multimodal deep learning autoencoder strategy, in summary, has outperformed conventional machine learning and deep learning techniques in cloud security. The technique detects complex patterns and abnormalities that may elude single-modality approaches, offering a full understanding of cloud security concerns through the fusion of many data sources and modalities. Because of its automatic feature learning, robustness against noisy data, and anomaly detection capabilities, the technique is a good fit for practical cloud security applications. The outcomes show how well the method works to identify a wide range of threats and anomalies, such as malware, anomalies, and threat classification, all with excellent accuracy. This discovery is important because it has the potential to completely transform cloud security by offering a strong and reliable defense against abnormalities and sophisticated attackers. This technique is appropriate for large-scale cloud security installations due to its scalability and ease of parallelization. Organizations can improve their threat detection capacities, lower the risk of cyberattacks, and safeguard critical data by incorporating this strategy into cloud security solutions.

In conclusion, our study has shown how successful a multi-modal deep learning autoencoder strategy is for cloud security, providing a viable way to identify abnormalities and sophisticated threats instantly. The method appeals to both researchers and practitioners of cloud security due to its automatic feature learning, robustness to noise, anomaly detection, and modality fusion. Future research directions for MMDLEA include exploring its application in edge AI and IoT security, as well as integrating Explainable AI (XAI) techniques to increase transparency and trust. Additionally, investigating MMDLEA's robustness against adversarial attacks and developing techniques to detect and mitigate such attacks is crucial. This will further enhance MMDLEA's capabilities and address emerging challenges in cloud security.

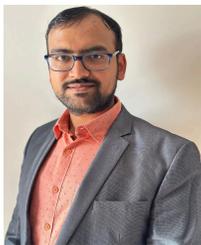

Aamiruddin Syed is a seasoned Security Professional with nearly a decade of experience in cybersecurity, specializing in DevSecOps, cloud security, and supply chain protection. He holds dual master's degrees in Cybersecurity and has authored book titled "Supply Chain Software Security". His research interests include integrating security automation into CI/CD pipelines, internal penetration testing, and enhancing security within critical industries like manufacturing. A recognized speaker at DEFCON and BlackHat, Aamiruddin actively contributes to the field through open-source projects and thought leadership, including his podcast "CyberGPT Pulse." He holds multiple industry certifications, including GIAC GCSA and CISA.